\documentclass[10pt,aps,prl,twocolumn,preprint,nofootinbib]{revtex4}

\usepackage{amsmath}    % need for subequations
\usepackage{mathtools}
\usepackage{graphicx}   % for figures
\usepackage{color}
\usepackage{verbatim}

\usepackage{cancel}

\begin{document}

\title{Gauge Invariance and Spontaneous Symmetry Breaking in Two-Gap Superconductors}
\author{Miguel C N Fiolhais}
\email{miguel.fiolhais@cern.ch}   %optional
\affiliation{Department of Physics, City College of the City University of New York, 160 Convent Avenue, New York 10031, NY, USA}
\affiliation{LIP, Department of Physics, University of Coimbra, 3004-516 Coimbra, Portugal}
\author{Joseph L Birman}
\email{birman@sci.ccny.cuny.edu}   %optional
\affiliation{Department of Physics, City College of the City University of New York, 160 Convent Avenue, New York 10031, NY, USA}
\date{\today}
%~\\[-1em]
\begin{abstract}

The gauge symmetry of the Ginzburg-Landau theory for two-gap superconductors is analyzed in this letter. We argue that the existence of two different phases, associated with the two independent scalar Higgs fields, explicitly breaks the gauge symmetry of the Ginzburg-Landau hamiltonian, unless a new additional vector field is included. Furthermore, the interference term, or Josephson coupling, holding a direct dependence with the phase difference, also explicitly breaks down the gauge symmetry. We show that a solution for the problem is achieved by adding an additional kinetic coupling term between the two vector fields, which generates the desired terms through a spontaneous symmetry breaking mechanism. Finally, the electrodynamics of the system is also presented in terms of the supercurrents inside the superconducting region.
\\
\\
The following paper is published in Physics Letters A: http://www.sciencedirect.com/science/article/pii/S0375960114006859

%The resulting two-dimensional quartic potential is studied after the spontaneous symmetry breaking, where two possible minima appear, leading to the possible co-existance of two Meissner states inside the superconductor. %A more complete model is presented in this letter for such class of superconductors, containing additional higher-order interference terms to preserve gauge invariance, and allowing phase differences between the two scalar fields nonetheless.

\end{abstract}

\maketitle

\section{Introduction}

The discovery of superconductivity in magnesium diboride MgB$_2$ in 2001~\cite{budko}, revealed for the first time the existence of two gaps in superconductors. The phenomenological description of such systems can be achieved through an extended multi-component Ginzburg-Landau theory~\cite{ginzburg}, analogue of other extensions also found in liquid metallic hydrogen~\cite{yokoya}, or in high-energy physics, in two-Higgs-doublet models (2HDM)~\cite{branco}. 

The phenomenology associated with such systems becomes particularly interesting, especially in superconductors, when they exhibit a mixture of properties from type-I and type-II superconductivity, also known as type-1.5~\cite{babaev,carlstrom,silaev,babaev2,moshchalkov}. The theoretical macroscopic description of these class of superconductors has been under study during the past decade, with several phenomenological predictions derived from the multi-component Ginzburg-Landau theory~\cite{gurevich1,gurevich2}. Moreover, the multi-band Ginzburg-Landau parameters have also been microscopically derived from the BCS hamiltonian near the phase transition~\cite{shanenko,orlova,vagov}.

In this letter, the gauge invariance of multi-band superconductors is studied and reviewed, leading to the conclusion that $n$ vector fields are necessary to preserve a local gauge symmetry of a superconductor with $n$ complex scalar order parameters, holding independent phases. It is also shown that the Josephson interaction terms, which also explicitly violate the local gauge invariance, can be generated through a spontaneous symmetry breaking mechanism from the coupling between the different gauge vector fields. The total supercurrent associated with these gauge fields is also derived in this framework, leading to the physical interpretation of the global magnetic field inside the superconducting region. The possible vortex solutions for these supercurrents are also qualitatively analyzed in the context of previous work with similar models.

\section{Ginzburg-Landau theory and Gauge Invariance}

The Ginzburg-Landau hamiltonian for a two-band superconductor at zero Kelvin is usually written in its most simple form as,
\begin{eqnarray}
 \mathcal{H} & = &  \frac{1}{2m_1} \left ( {\mathbf{\mathbf{\nabla}}} + 2 i e \mathbf{A} \right )\psi_1^* \left (\mathbf{\nabla} - 2 i e \mathbf{A} \right ) \psi_1  \nonumber \\ 
 & + &  \frac{1}{2m_2} \left ( {\mathbf{\nabla}} + 2 i e \mathbf{A} \right )\psi_2^* \left ({\mathbf{\nabla}} - 2 i e \mathbf{A} \right ) \psi_2\nonumber \\
& + & \alpha_1 |\psi_1|^2 + \alpha_2 |\psi_2|^2 + \beta_1 |\psi_1|^4 + \beta_2 |\psi_2|^4 \nonumber \\ 
& + & \frac{1}{2\mu_0} \left ( {\mathbf{\nabla}} \times \mathbf{A} \right )^2 \, ,
\label{hamiltonian1}
\end{eqnarray}
\begin{comment}
\begin{eqnarray}
 \mathcal{H} & = &  \frac{1}{2m} \left ( {\mathbf{\nabla}} + 2 i e \mathbf{A} \right )\psi_1^* \left ({\mathbf{\nabla}} - 2 i e \mathbf{A} \right ) \psi_1  \nonumber \\ 
 & + &  \frac{1}{2m} \left ( {\mathbf{\nabla}} + 2 i e \mathbf{A} \right )\psi_2^* \left ({\mathbf{\nabla}} - 2 i e \mathbf{A} \right ) \psi_2\nonumber \\
 & + &  \frac{\gamma}{2m} \left ( {\mathbf{\nabla}} + 2 i e \mathbf{A} \right )\psi_2^* \left ({\mathbf{\nabla}} - 2 i e \mathbf{A} \right ) \psi_1\nonumber \\
 & + &  \frac{\gamma}{2m} \left ( {\mathbf{\nabla}} + 2 i e \mathbf{A} \right )\psi_1^* \left ({\mathbf{\nabla}} - 2 i e \mathbf{A} \right ) \psi_2\nonumber \\
& + & \alpha_1 |\psi_1|^2 + \alpha_2 |\psi_2|^2 -\eta(\psi_1^*\psi_2+\psi_1 \psi_2^*) \nonumber \\
& + & \beta_1 |\psi_1|^4 + \beta_2 |\psi_2|^4 + \nu |\psi_1|^2|\psi_2|^2 \nonumber \\
&+& \frac{1}{2\mu_0} \left ( {\mathbf{\nabla}} \times \mathbf{A} \right )^2 \, ,
\label{hamiltonian1}
\end{eqnarray}
\end{comment}
with the order parameter $\psi_i(x) = \rho_i (x) \textrm{e}^{i e \theta_i (x)}$, where $\rho_i(x)$ and $\theta_i (x)$ are real fields. The vector field is represented by $\mathbf{A}$, $2e$ is the electric charge of the Cooper pairs~\cite{cooper,bardeen,bardeen2,suhl}, and the real constants $\alpha_i$, and $\beta_i$ are the arbitrary parameters of the theory. The bare masses of the condensates pairs, $m_1$ and $m_2$, are assumed to be different to comprise systems like liquid metallic hydrogen or liquid metallic deuterium, where the mass of the deuteron is three orders of magnitude larger than the Cooper pairs'~\cite{zou,garaud}.

The existence of different phases, associated with the two scalar fields, immediately raises a problem related to the gauge invariance of the theory, as discussed in~\cite{yanagisawa}. As both fields couple to the same vector field, the $U(1)$ local gauge invariance cannot be preserved if both scalar fields have different phases, with the vector field transforming as
\begin{eqnarray}
\psi_i (x) & \rightarrow & \psi_i ' (x) \equiv \textrm{e}^{2ie\theta_i (x)} \psi_i (x) \\
\mathbf{A} (x) & \rightarrow & \mathbf{A}' (x) \equiv \mathbf{A} (x) - \frac{1}{2e} \mathbf{\nabla} \vartheta (x) \, .
\end{eqnarray}
To build a local gauge invariant theory with two independent scalar fields, a $U(1) \times U(1)$ symmetry is necessary, implying the existence of two vector fields inside the superconducting region. As such, the hamiltonian for a general non-interacting multi-band system can be written as
\begin{eqnarray}
 \mathcal{H} & = &  \sum_{i=1}^n  \frac{1}{2m_i} \left ( {\mathbf{\nabla}} + 2 i e \mathbf{A}_i \right )\psi_i^* \left ({\mathbf{\nabla}} - 2 i e \mathbf{A}_i \right ) \psi_i  \nonumber \\ 
& + & \alpha_i |\psi_i|^2 + \beta_i |\psi_i|^4 + \frac{1}{2\mu_0} \left ( {\mathbf{\nabla}} \times \mathbf{A}_i \right )^2  \, ,
\label{hamiltonian2}
\end{eqnarray}
with the fields transforming as
\begin{eqnarray}
\psi_i (x) & \rightarrow & \psi_i ' (x) \equiv \textrm{e}^{2ie\theta_i (x)} \psi_i (x) \\
\mathbf{A}_i (x) & \rightarrow & \mathbf{A}_i' (x) \equiv \mathbf{A}_i (x) - \frac{1}{2e} \mathbf{\nabla} \theta_i (x) \, ,
\end{eqnarray}
under the $i$th $U(1)$ local gauge transformation. In this case, the  vector fields do not couple with each other and neither do the complex scalar fields, which means that the electromagnetic field independently interacts with the different condensates. By applying the usual Higgs mechanism to spontaneously break the local gauge symmetry~\cite{anderson,englert,higgs,guralnik,atlas,cms}, the independent vector fields acquire a mass~\cite{kleinert2,essen,fiolhaisessen,fiolhais2,fiolhais3,fiolhaiskleinert}, giving rise to an overlap of Meissner or vortex states. The gauge invariant potential, in the case both quadratic parameters are negative, 
\begin{eqnarray}
V(\rho) & = &  \alpha_1 \phi_1^2 + \alpha_2 \phi_2^2 + \beta_1 \phi_1^4 + \beta_2 \phi_2^4 \, .
\label{potential1}
\end{eqnarray}
has a non-degenerate minimum value after the spontaneous symmetry breaking,
\begin{eqnarray}
(\rho_1, \rho_2) & = & \left (\phantom{-}\frac{1}{\sqrt{2}}\sqrt{-\frac{\alpha_1}{\beta_1}} , \phantom{-} \frac{1}{\sqrt{2}}\sqrt{-\frac{\alpha_2}{\beta_2}}\phantom{-} \right ) \, .
\label{potential3}
\end{eqnarray}
In other words, the photons inside the superconducting region may acquire different effective masses, depending on which boson condensate they interact with. This is, in fact, not a surprising result, as several condensed matter systems comprise overlaps of several modes with different effective masses.
It should be noted, however, that this model does not comprise a change of nature of the electromagnetic field as we know it inside the multi-band superconducting region. Instead, the two gauge fields appear as a macroscopic description of the interaction between the electromagnetic field and the two condensates, just as the massive photons in traditional Ginzburg-Landau theory.

\section{Josephson Coupling and Spontaneous Symmetry Breaking}

This previous model is not very realistic, however, in the sense that it assumes no interaction at all between the condensates. To parameterize such interactions, usually a Josephson interference term is added to the hamiltonian~\cite{gurevich},
\begin{equation}
 \mathcal{H}^{\textrm{jph}} = -\eta \rho_1 \rho_2 \cos (\theta_2-\theta_1) \, ,
\end{equation}
with a real coupling constant $\eta$. Nevertheless, the same problem arises as such interference term directly depends on the phase difference and violates the gauge freedom. In other words, the different choice of the complex phases leads to different outcomes, \emph{i.e.} non-physical results.

To overcome this difficulty, another gauge invariant interference kinetic term can be included in the hamiltonian, namely
\begin{equation}
 \mathcal{H}^{\textrm{int}} = \gamma \left ( {\mathbf{\nabla}} + 2 i e \mathbf{A}_2 \right )\psi_2^* \left ({\mathbf{\nabla}} - 2 i e \mathbf{A}_1 \right ) \psi_1 + h.c.\, ,
\end{equation}
where $\gamma$ gives the strength of the interaction between the two massive photon fields. By expanding the complex scalar fields in terms of its real phases, a Josephson-like term arises:
\begin{equation}
 \mathcal{H}^{\textrm{jph}} = 4 e^2 \gamma ( \mathbf{\nabla} \theta_1 \cdot \mathbf{\nabla} \theta_2 ) \rho_1 \rho_2 \cos (\theta_2-\theta_1) \, .
\end{equation}
Since the real phases are arbitrary due to the local gauge freedom, one can spontaneously break the symmetry and choose the appropriate phase functions values which provide the desired phenomenological description of the problem. In this case,
\begin{equation}
\mathbf{\nabla} \theta_1 \cdot \mathbf{\nabla} \theta_2 = -\frac{\eta}{4 e^2 \gamma} \, ,
\end{equation}
and the Josephson coupling term is sucessfully generated through a spontaneous symmetry breaking mechanism. 

Finally, the super-currents can also be derived in this model, yielding
\begin{eqnarray}
\mathbf{j}_1 & = & - \frac{4e^2}{m_1} |\psi_1|^2 \mathbf{A}_1 \nonumber \\
&& - 4e^2 \gamma ( \psi_2^* \psi_1 + \psi_1^* \psi_2 ) \mathbf{A}_2 \nonumber \\
&& + \frac{2ie}{m_1} (\psi_1 \mathbf{\nabla} \psi_1^*+\psi_1^* \mathbf{\nabla} \psi_1) \nonumber \\
&& + 2ie\gamma (\psi_2 \mathbf{\nabla} \psi_1^*+\psi_1^* \mathbf{\nabla} \psi_2) \, , 
\end{eqnarray}
and,
\begin{eqnarray}
\mathbf{j}_2 & = & - \frac{4e^2}{m_2} |\psi_2|^2 \mathbf{A}_2 \nonumber \\
&& - 4e^2 \gamma ( \psi_1^* \psi_2 + \psi_2^* \psi_1 ) \mathbf{A}_1 \nonumber \\
&& + \frac{2ie }{m_2} (\psi_2 \mathbf{\nabla} \psi_2^*+\psi_2^* \mathbf{\nabla} \psi_2) \nonumber \\
&& + 2ie\gamma (\psi_1 \mathbf{\nabla} \psi_2^*+\psi_2^* \mathbf{\nabla} \psi_1)  \, ,
\end{eqnarray}
where the dependency with the phase difference becomes, once more, explicit due to the interference terms between the vector fields. It should be stressed, however, that these super-currents are, in fact, locally gauge invariant, and therefore, physical observables that can be experimentally measured, despite the presence of terms with explicit phase differences. An elucidative exercise can be performed by applying the previous transformations to these super-currents, preserving the local gauge invariance.

The total current in a two-gap superconductor is, therefore,
\begin{eqnarray}
\mathbf{j} & = & \mathbf{j}_1+\mathbf{j}_2 \, ,
% & = & \mathbf{\nabla} \times \mathbf{\nabla} \times (\mathbf{A}_1+\mathbf{A}_2) \nonumber  \, ,
\end{eqnarray}
and can be associated with the global magnetic field inside the superconducting region,
\begin{eqnarray}
\mathbf{B} & = & \mathbf{\nabla} \times (\mathbf{A}_1+\mathbf{A}_2) \, .
% & = & \mathbf{\nabla} \times \mathbf{\nabla} \times (\mathbf{A}_1+\mathbf{A}_2) \nonumber  \, ,
\end{eqnarray}
The final magnetic field results, therefore, from the combination of the two gauge vector fields, and encompasses several possible configurations, such as pure Meissner states, or mixtures of vortices states, depending on the properties of each condensate~\cite{kogan,brandt}. The coupling terms may also play an important role, resulting in tangled vortices solutions, whose study is beyond the scope of this letter.

\section{Conclusions}

It is, thus, clear that two vector fields must exist inside two-gap superconductors, each corresponding to different massive photons, to fully preserve the local $U(1) \times U(1)$ gauge symmetry. In fact, the interference terms between the two massive photon fields also give rise to the forbidden Josephson couplings between scalars, through a spontaneous symmetry breaking mechanism.
These results may hopefully lead to a change of paradigm when dealing with such systems, centering the construction of theories around the local gauge symmetry groups, and by attempting to derive any desired anomalous terms through spontaneous symmetry breaking mechanisms.

\section{Acknowledgments}

The work of M.C.N.~Fiolhais was supported by LIP-Laborat\'orio de Instrumenta\c c\~ao e F\'isica Experimental de Part\'iculas, Portugal.

\end{document}